\documentstyle[epsfj]{article}
\begin{document}
\begin{titlepage}
\begin{flushright}
TIT/HEP-298/COSMO-57 \\
August \ 1995 \hspace{1.7cm} \\
\end{flushright}
\vskip 3cm
\begin{center}
\Large{Anderson Localization in the Sky \\
and Cosmological Magnetic Field}
\end{center}
\vskip 2cm
\begin{center}
Akio HOSOYA\footnote{E-mail address: ahosoya@th.phys.titech.ac.jp}
and Shiho KOBAYASHI\footnote{E-mail address: kobayasi@th.phys.titech.ac.jp}\\
\vskip 0.5cm
Department of Physics, Tokyo Institute of Technology\\
Oh-Okayama, Meguro-ku,Tokyo 152, Japan
\end{center}
\vskip 3cm
\begin{abstract}
 We discuss the Anderson localization of electromagnetic fields
in the fluctuating plasma induced by the gravitational density perturbation
before the recombination time of the Universe. Randomly distributed localized
coherent electromagnetic fields emerge in the thermal equilibrium
before the recombination time. We argue that the localized coherent
electric fields eventually produce cosmological magnetic fields
after the decoupling time.
\end{abstract}
\vskip  1cm
PACS number(s): 52.40.Db,71.55.Jv,98.60.Jk,98.80.Dr
\vskip 0.5 cm
\end{titlepage}

\section{Introduction}
It has long been suspected that the presently observed galactic and cluster
magnetic fields ($10^{-6}gauss$)\cite{ASSE}
(and possibly the intercluster magnetic field $\approx 10^{-9}gauss$,
existence of which
is under disputes \cite{ASSE}\cite{SFK}\cite{TF}.) has a cosmological
origin. There are two competing theories
for the galactic magnetic field: the primeval scenario and the dynamo
theory. The former explains the galactic magnetic field as a
consequence of amplification of a
seed magnetic field by protogalaxy collapse.
On the other hand, in the more popular galactic dynamo theory
\cite{SF}\cite{PAR}, the magnetic field has been
exponentially amplified from a tiny seed magnetic field by differential
galactic rotation and turbulence. One of the weak points of the
primeval magnetic field scenario has been the lack of its viable explanation.

Nevertheless, an alternative theory of density irregularities on the basis of
the primeval magnetic field  \cite{WAS}\cite{KOR} is attractive, because it can
explain the magnitude of the density perturbation and the galactic
magnetic field simultaneously. Recently a direct estimate of magnetic
fields of young
galaxies ($z \approx 2$) has been performed by measuring the Faraday rotation
of quasar spectra through the Lyman-$\alpha$ absorption systems to
give $10^{-6} gauss$ \cite{WLO}\cite{AA}.
This large magnetic field of the galaxy in the early stage may revive
interest in
the primeval scenario. Further, even for the galactic dynamo theory
we need a seed
magnetic field, however small it is.

In this paper we would like to point out that before the recombination
time the cosmic plasma is strongly fluctuating
at large scales due to the gravitational density perturbation.
The plasma density fluctuation plays a role of randomly distributed scatterers
for the electromagnetic fields and make it impossible
for the electromagnetic fields
to propagate if the wave length is longer than the elastic mean
free path of electromagnetic
wave by the scatterers. The localized modes are thermally
excited and the electric field is more or less aligned in the localization
region. The localized electric field induces
a coherent current and then the magnetic field $\approx
10^{-26}\;\;gauss$ at the time of decoupling of photons from matter.

The localization phenomenon by random fields was first proposed
by Anderson \cite{AND} to explain the behavior of the electric
resistivity of metals in the presence of
impurities and the theories were developed by himself and
his collaborators \cite{AALR}\cite{THOU}.
The localization phenomena are observed in various physical systems,
e.g. in the propagation of light waves \cite{JOHN} and of sound
waves \cite{HM}.
Probably the first paper which studied the possible localization
of electromagnetic
field in plasma is the one by Escande and Souillard \cite{ES}.
A preliminary experiment was performed
for the longitudinal electron plasma wave by Doveil et al. \cite{DVT}.
(We shall adopt the localization theory in plasma \cite{ES}
to the cosmological plasma in the expanding Universe.)

Roughly speaking, the Anderson localization phenomenon
occurs because of the destructive interference of the incident
wave and the scattered waves from
the randomly distributed sources.
It is known that the localization occurs if the strength of the randomness is
larger than a critical value. For a sufficiently large random
potential, only the modes
of the wave length shorter than the mean free path can
propagate, while the other modes are localized.

We are going to argue that this localization
phenomenon should occur in the matter
dominated era in the standard
cosmology. The resultant localized magnetic field may explain
the origin of the galactic magnetic field
and possibly a structure formation in the early stage \cite{WAS}\cite{KOR}.

This paper is organized as follows. After briefly reviewing
the electromagnetism
in the cosmological plasma in \S2, we will derive the basic
equation for the Anderson
localization of electromagnetic fields in the standard big
bang model of the Universe
in \S3. In \S4 and \S5, we will estimate the mobility edge,
the localization length and the number of localized states.
We will describe a scenario which tells how the localized
coherent electromagnetic fields emerge before the recombination
time in \S6 and give a rough estimate
of the cosmological magnetic field at the decoupling time.
The final section \S7 is devoted to summary and discussions.

\section{The Plasma Equations in the Friedmann-Robertson-Walker Universe}

Here we will first review the plasma equations
in the Friedmann-Robertson-Walker background.

Let us recall that the relativistic expression for the Maxwell
equation is given by
 \begin{equation}
F^{\alpha \beta}|_{\beta}=j^{\alpha},
\end{equation}
where the electromagnetic field strength
$F_{\alpha \beta}=\partial_{\alpha}A_{\beta}
-\partial_{\beta}A_{\alpha}$ with $A_{\alpha}$ being the vector potential.
Here the current $j^{\alpha}=enu^{\alpha}$ with $u^{\alpha}$
being the four velocity
of charged particles(electrons) with $u_{\alpha}u^{\alpha}=-1$
and $n$ the density.
The velocity $u^{\alpha}$ satisfies the Lorentz force equation:
 \begin{equation}
mu^{\alpha}u_{\lambda}|_{\alpha}= F_{\mu\alpha}eu^{\alpha}.
\end{equation}
For simplicity, consider the spatially flat
Friedmann-Robertson-Walker Universe with
the metric
\begin{equation}
ds^2=a(\lambda)^2[-d\lambda^2+d{\bf x}^2].
\end{equation}
Here $\lambda$ is the conformal time defined by
$\lambda=\int^t{dt \over a}$.
We assume that the charged particles are non-relativistic.
So $u^{\alpha}=(a^{-1},v^i)$
with $v^i$ being small compared with unity.
Then the Maxwell equation reads
\begin{equation}
{\eta}^{\alpha \beta}\partial_{\beta}F_{i \alpha}=ena^2v_{i}.\label{maxwell}
\end{equation}
with ${\eta}^{\alpha \beta}$ being the Lorentzian metric
and the particle equation is approximately given by
\begin{equation}
m{d v_i \over d \lambda}\approx eF_{i0},\label{lorentz}
\end{equation}
where $m$ is the electron mass and $e$ is its charge.
{}From the conservation of the current $j^{\mu}|_{\mu}=0$, we see that $na^3$
is
 approximately constant in our non-relativistic case
before the recombination time.
\section{Random Potential by Fluctuating Plasma Density}
Combining the equations (\ref{maxwell}) and (\ref{lorentz}),
we can derive a general relativistic equation for the vector
potential ${\bf A}$  in terms of
the conformal comoving coordinates $(\lambda,{\bf x})$ by
\begin{equation}
-{\partial^2 {\bf A} \over \partial \lambda^2}+
{\partial^2 {\bf A} \over \partial{\bf x}^2}
  ={ e^2na^2 \over m}{\bf A}.
\end{equation}
Split the plasma density $n$ into the homogeneous part $\overline{n}$
and the space dependent random part $\delta n$;
\begin{equation}
n = \overline{n}+\delta n.
\end{equation}
Then the right hand side of the plasma equation becomes
\begin{equation}
 [{\overline{\omega_p^{rec}}^2 \over a} +{\overline{\omega_p^{rec}}^2 \over a}
 {\delta n \over \overline{n}}]{\bf A},
\end{equation}
where $\overline{\omega_p^{rec}}=\sqrt{{e^2 n_{rec} \over m_e}}$.

Here we have used the conservation law:$\overline{n}=n_{rec}/a^3$
before the recombination time,
with the scale factor $a$ normalized at the recombination time.
$n_{rec}$ is the electron density
(therefore the baryon density) at the recombination time.

The first term in the bracket is spatially constant but a decreasing
function of the
conformal time $\lambda$. The second term represents the fluctuating random
environment induced by the baryonic density perturbation
${\delta\rho_b \over \rho_b}$
which is equal to ${\delta n \over \overline{n}}$ in our non-relativistic case.
Knowing the standard theory of linear perturbation in cosmology we see that
${\delta n \over \overline{n}}={\delta\rho_b \over \rho_b} \propto a$
in the matter dominated era,
so that the second term is independent of the conformal time in that era.
We write the random potential as
$V({\bf x})\equiv {\overline{\omega_p^{rec}}^2 \over a}{\delta n \over
\overline{n}}$, which is a function of comoving coordinates ${\bf x}$ only.
So far we have discussed the period after the equal time and before the
recombination time and found that the random potential
induced by the density perturbation
is constant in time. As we shall see this is important for the stability of the
Anderson localization. After the recombination time,
the density of free electrons sharply
decreases so that the random potential changes
as rapidly as the other terms in the
Maxwell equations so that the localization breaks down
after the recombination time.
We will discuss this localization and de-localization
phenomena in more detail in \S6.

For the matter dominated universe, the scale factor is given by
$a(t) =({t \over t_{rec}})^{2/3}=({\lambda \over 3t_{rec}})^2 $
with $\lambda =\int{dt \over a(t)}=t_{rec}^{2/3}3t^{1/3}$.
Putting them all together we reduce the Maxwell equations to the following
wave equation as
\begin{equation}
-{\partial^2 {\bf A} \over \partial \lambda^2}+
{\partial^2 {\bf A} \over \partial{\bf x}^2}
= [Q^2/\lambda^2+V({\bf x})]{\bf A}.
\end{equation}
Here $Q^2=9t_{rec}^2\overline{\omega_p^{rec}}^2$.

We expand the vector potential ${\bf A}$ in terms of the  complete orthonormal
set $\{\psi_n\}$ as
\begin{equation}
{\bf A}=\sum {\bf a}_nf_n(\lambda)\psi_n({\bf x})+c.c.
\end{equation}
with ${\bf a}_n$ being arbitrary coefficients.
Here $\psi_{n}({\bf x})$ is a normalized solution of the eigenvalue equation:
\begin{equation}
[-\Delta+V({\bf x})]\psi_n({\bf x})=E_n^2\psi_n({\bf x})\label{sch}.
\end{equation}
The wave equation reduces to
\begin{eqnarray}
&{d^2 f_n \over d \lambda^2}+{Q^2 \over \lambda^2}f_n +E_n^2 f_n \cr
&\approx{d^2 f_n \over d \lambda^2}+{Q^2 \over \lambda^2}f_n =0,
\end{eqnarray}
with the normalization condition:
$i(f_n^{*}\partial_{\lambda}f_n-f_n\partial_{\lambda}f_n^{*})=1$.

Localization means that a number of modes
$\{\psi_n, n\leq n_0\}$ are bound states
exponentially falling off at infinity. The marginal state $ n_0$ corresponds to
the so-called mobility edge as we shall explain in the next section.

\section{Mobility Edge and Mean Free Path}
Throughout this paper
we only consider the adiabatic baryonic density perturbation for
simplicity, the spectrum of which is roughly like the
picture below. To simplify our discussion we assume
that the spectrum is sharply peaked
around the Jeans length $l_c$ at the equal
time \footnote{We are fully aware that
inclusion of the "scale invariant spectrum"
for scales larger than $l_c$ produces a hierarchy of structures.
We would like to postpone
the investigation in this direction to future works.}
and the distribution of perturbation is Gaussian.
\begin{figure}
\begin{center}
\epsfile{file=fig1.eps,width=8cm}
\end{center}
\caption{The spectrum of the adiabatic baryonic perturbation}
\end{figure}
To understand the Anderson localization in the Universe
before the recombination time, we
have to evaluate the mobility edge of the wave number,
which is roughly the inverse of
the mean free path $l_{mf}$ of the electromagnetic wave by the
random potential.
Consider a Schr\"{o}dinger equation for the spatial part
of the electromagnetic field:
\begin{equation}
[-\Delta+V({\bf x})]\psi=E^2\psi,
\end{equation}
with ${\bf x}$ being the comoving coordinates.
Recall that the random potential
$V({\bf x})=\overline{\omega_p^{rec}}^2{\delta n \over \overline{n}}$
does not depend on the conformal time in the matter dominated era.
The characteristic length scale $l_c$ of the density perturbation
is given by the Jeans length at the equal time\footnote{The equal
time $t_{eq}$ is the time when the matter and radiation energy
density are equal. This value depends on the value of the present
baryon density. We take ${a(t_{rec})
\over  a(t_{eq})}= ({t_{rec} \over t_{eq}})^{2/3}=6$ as a
typical value for $\Omega_{0}=0.2$.}, which is roughly the same
as the horizon scale at that time. After that time the
length scale $l_c$ develops as
$\propto a$. This gives $l_c\approx10^{23}cm$ at the
recombination time, while the horizon is $\approx 2\times10^{23}cm$.

Then the mean free path is given by
 \begin{equation}
l_{mf}={1 \over n_c\sigma},
 \end{equation}
where $\sigma$ is the elastic scattering cross section of
the electromagnetic wave by a single potential and
$n_c\approx{3 \over 4\pi} l_c^{-3}$ is the density of
"impurities".
The elastic cross section is estimated roughly as $\approx 4\pi l_c^2$ by the
standard wave mechanics, because the
"energy" is low and is roughly given by $E^2 l_c^2\approx 1$,
while the potential is
effectively gigantic $V\times l_c^2\approx
\overline{\omega_p}^2{\delta n \over n}l_c^2\approx 5\times10^{32}$.
(Here we have taken
${\delta n \over n}\approx10^{-5}$ because the assumed  baryonic perturbation
can be comparable
to the temperature fluctuation of CMB observed by COBE \cite{COBE}.
However, the localization length
and other results in the present work are insensitive to
the magnitude of the density perturbation.)
Therefore the localization should occur, since the critical value of $Vl_c^2$
will be of order one.
Then the mean free path is given by
\begin{equation}
l_{mf}\approx l_c/3,
\end{equation}
and the mobility edge is $E^{*}\approx {12\pi \over l_c}$,
which is consistent with our low energy picture.
\begin{figure}
\begin{center}
\epsfile{file=fig2.eps,width=8cm}
\end{center}
\caption{The characteristic length $l_c$, mean free path $l_{mf}$
and localization length $\xi$. }
\end{figure}

\section{Localization Length}
Fortunately in our present case, the randomness is huge so that an estimate of
the localization length is available \cite{AM}.
We have
 \begin{equation}
\xi=l_{c}/\log(|V|/E^2),
 \end{equation}
where $l_{c}$ is the minimum length scale
in which range the potential can be considered
as a constant (the same notation as the one by Escande and
Souillard \cite{ES}). For the adiabatic baryonic
perturbation, it will be reasonable to take the Jeans length
at the equal time $t_{eq}$
as the small scale cut-off $l_{c}$, which is
$\approx 10^{23} cm$ at the recombination time.
In our present case, $|V|/E^2\approx5\times10^{32}$ so that
$\xi\approx 10^{21} cm$ at the recombination time.

It is easy to estimate the number of localized
states $N$ by equating it with the
number of states below the mobility edge $E^{*}=2\pi l_{mf}^{-1}$.
Namely,
 \begin{equation}
{V \over (2\pi)^3}2\times\int^{E^{*}}d^3k=N.
 \end{equation}
  We find that the density of the localized states is given by
  \begin{equation}
N/V={8\pi \over 3}l_{mf}^{-3}.
 \end{equation}
 There will be around $2000$ lumps of electromagnetic fields
inside the horizon at the recombination time.
We will not discuss its direct observational implications
in the present work, because
there are uncertainties in the estimate of
the mobility edge so that we should not
take the predicted number of the lumps too seriously.
\section{Cosmological Magnetic Field}
The localized modes in the previous sections are
excited in the thermal equilibrium
 after the equal time and before the recombination time.
We estimate the magnitude of
 the thermally excited localized electric fields.
 The localized electric field ${\bf E}_{loc}$ is expanded as
\begin{equation}
{\bf E}_{loc}= -\sum ^{n_0}{\bf a}_n\dot{f_n}(\lambda)\psi_n({\bf x})+c.c.
\end{equation}
so that the thermal average of ${\bf E}_{loc}^2$ is
\begin{eqnarray}
\overline{{\bf E}_{loc}^2} &= \sum^{n_0}\overline{{\bf a}_n^{*}{\bf a}_n}
|\dot{f_n}|^2|\psi_n|^2\cr
               &=\sum^{n_0}{1 \over e^{Q/\lambda T}-1}|\dot{f_n}|^2|\psi_n|^2.
\end{eqnarray}
Since $Q/\lambda T<<1$, $|\dot{f_n}|^2\approx Q/\lambda$ and
$|\psi_n|^2\approx \xi^{-3}$,
we obtain
\begin{equation}
\overline{{\bf E}_{loc}^2}\approx T/\xi^3,
\end{equation}
in a localization region of the size $\xi$. Here we have taken
into account the contribution
from a single localized mode only, because the peaks of the other
localized modes are
randomly distributed and
will be located outside of the region under consideration.
The estimate of $\overline{{\bf E}_{loc}^2}$ can also be
derived by the equal partition law.
Note that the propagating modes($n>n_0
$) contribute to the Stefan-Boltzmann law.

At this stage we realize that approximately $T/\omega_p$ photons occupies
a localized state. As far as $T/\omega_p>>1$, a coherent localized
state of the size $\xi$ is a good picture so that we can approximately
describe it by a classical
field:
\begin{equation}
{\bf E}_{loc}= -\sum ^{n_0}\sqrt{T/\omega_p}\dot{f_n}(\lambda)
{\bf e}_n({\bf x})\psi_n({\bf x})+c.c.,
\end{equation}
where ${\bf e}_n({\bf x})$ is the polarization vector,
which is perpendicular to the
gradient of the wave function $\psi_n({\bf x})$.
Therefore, in a localization domain of the size $\xi$
in the cosmic plasma, we have aligned electric
and magnetic fields, which are perpendicular to each other and
oscillate almost at the plasma frequency.

One might worry about the damping of the magnetic structure of the size $\xi$
by photon diffusion. However, the diffusion does not occur to the
localized modes.
This is one of the virtues of the Anderson localization mechanism.

Since the localization is a delicate interference phenomenon, the
random potential $V({\bf x})$ should vary slowly so that the motion of wave
packet is not disturbed \cite{ES}. This condition is met  before
the recombination time,
which we have discussed so far.  After that time we have to
take into account the
recombination process in our equations for the electromagnetic fields.
The number of free electrons sharply decreases, and the random
potential changes in time.

After recombination time the packets of the localized coherent
electromagnetic fields
begin to diffuse. We will see the time evolution of the packet
by solving the Vlasov
equation for the electron distribution and the Maxwell equations
for the electromagnetic fields.
\begin{figure}
\begin{center}
\epsfile{file=fig3.eps,width=8cm}
\end{center}
\caption{The diffusion of the localized electromagnetic field}
\end{figure}
For our weak field we can linearize the  Vlasov equation as
\begin{equation}
\frac{\partial f_1}{\partial t}+ {\bf v}\cdot\frac{\partial
  f_1}{\partial {\bf x}}+\frac{e}{m}{\bf E}\cdot
\frac{\partial f_0}{\partial {\bf v}}=0,
\end{equation}
where $f_1(t,{\bf x},{\bf v})$ is a small correction to the equilibrium Maxwell
distribution function $f_0({\bf v})$.
We assume that the initial distribution function
is the Maxwell distribution around the coherent electron velocity
${\bf u}=\frac{e}{m \omega_p}{\bf E}$,
\begin{equation}
f(t,{\bf x},{\bf v})=f_0({\bf v}-{\bf u}({\bf x}))   \ \ \ \ \ \  t <  t_{rec}.
\end{equation}
To solve the transport equation and the Maxwell equations,
the standard method is to use the Laplace transformation,
\begin{equation}
\widetilde{f_{1}}(\omega,{\bf v})=\int_0^\infty dt e^{i\omega t}
f_{1}(t,{\bf v}).
\end{equation}
In the Laplace transformed space we can easily obtain an electric
current density
$\widetilde{\bf j}=e\int d^3v{\bf v}\widetilde{f_{1}}$,
and then evaluate the Laplace inverse transform.
Since the high frequency parts of the current density are averaged out,
we evaluate only a part which originates
from ``the $\omega = {\bf kv}$ pole'' with ${\bf k}$
being the wave vector in the Fourier transformed space
(see \S34 in \cite{LP} for the detailed calculation ).
The result is
\begin{equation}
{\bf j}(t,{\bf x})\approx -\frac{e^2}{m}
\int d^3v f_0({\bf v}){\bf E}(0,{\bf x}-{\bf v}t).
\end{equation}
In order to roughly estimate the total current $I$, we may simply suppose
that the localized electromagnetic field has a Gaussian
form $E=(T/\xi^3)^{1/2}exp[-({\bf x}/\xi)^2]$ at the recombination time.
The induced magnetic $B$ is then given by
\begin{equation}
B\approx\frac{I}{2\pi \xi}\approx\frac{n r_0 \xi}{2 \omega_p}
\left(\frac{T}{\xi^3}\right)^{1/2}\approx 10^{-26}gauss,
\end{equation}
at the decoupling time with $r_0$ being the classical radius of
electron.

This linear picture holds until the ${\bf v}\times {\bf B}$ term
becomes important,
and after then the  magnetohydrodynamics stage will set in.
Since the localized region has a gigantic conductance,
the magnetic field created at the decoupling era stays
and will not dissipate to the Joule heat.

\section{Conclusion and Discussions}
In this paper we presented a new possible scenario which is intended
to explain the origin of cosmological magnetic field.

The baryonic density perturbation plays a role of random potential of the size
$10^{23}cm$ to the electromagnetic fields and triggers the Anderson
localization of the electric field before the recombination time.
A large number of low energy photons occupy
the localized states in the thermal equilibrium.
The resultant coherent localized  electric field is
aligned in the localization
 region. The localized electric field then induces
 a coherent current and  the magnetic field $\approx 10^{-26}\;\;gauss$ at the
 decoupling time.

Roughly, we expect 2000 magnetic lumps of the size $10^{21}cm$  inside
the horizon $2\times10^{23}cm$ at
that time. If the primeval magnetic field ($10^{-26}\,\,gauss$) at the
decoupling time is efficiently amplified by dynamo or other mechanism
in the later stage,
it may provide an explanation of the seed  magnetic field in the Universe.
 However, it is beyond the scope of the present paper to work out
the evolution of the magnetic fields in the long history of
the Universe after the
decoupling time. It is possible to argue that the resultant magnetic field
traps sufficient number of protons so that the gravitational
instability eventually creates a galaxy at the localization region.

In the present work, we assumed
that the large scale density perturbation is primordial and its magnitude
is chosen to be around ${\delta n \over \overline{n}}\approx
10^{-5}$ so as to be consistent with the COBE data on the anisotropy of CMB.
In this case, the large scale structure is primordial and the galaxy
scale one is its consequence by the Anderson localization.
As remarked in the text,
we only took into account a single scale $l_c$
for the primordial perturbation for
simplicity. The contribution from the larger scale to
the localization phenomenon
is interesting in the sense that it may account
for the hierarchy of cosmological
structures.

 It is well known that a density perturbation of a scale larger
than the Jeans length $l_J={c_s \over \sqrt{\rho G}}$ gravitationally
collapses against the thermal pressure
gradient and therefore the perturbation grows.
An analogous phenomenon exists for
the gravitational collapse of localized magnetic fields.
Correspondingly the magnetic
Jeans length is obtained by replacing the sound velocity $c_s$
in the expression of $l_{J}$ by the Alfv\'en velocity $c_a=B/\sqrt{\rho}$.
\begin{equation}
l_{mJ}={B \over \rho\sqrt{G}}
\end{equation}
(For more satisfactory derivation, see Kim et al.\cite{KOR}.)
Just after recombination time, $\rho\approx 10^{-20}g/cm^3$,
we obtain
 \begin{equation}
l_{mJ}\approx{B \over gauss}\times 10^{26}cm,
\end{equation}
which is comfortably less than the previously estimated localization length
$10^{21}cm$, if $B$ is less than the present galactic value $10^{-6}\,gauss$.
Therefore the gravity exceeds the magnetic pressure so that
the magnetic lumps will not blow up after the decoupling time.


\vskip 1 cm
\LARGE {\bf Acknowledgments} \par
\vskip 5 mm
\normalsize
We would like to thank Prof.M. Fujimoto for his nice talk
on galactic magnetic fields and for his useful advises.
The discussions with Dr. S. Adachi on the Anderson localization
and Prof. H. Morikawa for cosmological non-equilibrium processes
are especially helpful to us.
The comments by Prof. H. Ishihara have been valuable.
We also thank Prof. D.F. Escande for sending a material on the localization
in plasma and Prof. S. Kotani for informing us a valuable paper
in the reference \cite{AM}. One of the authors (A.H.) thanks Prof.
K. Shibata for opening up his eyes to the possible importance
of plasma physics in the early Universe \cite{KS}.

S.K. acknowledges the support by the Fellowships of
Japan Society for the Promotion of Science for Japanese Junior
Scientists.  This work is partially  supported by the Grants-in-Aid
for Scientific Research of the Ministry of Education, Science and
Culture of Japan (No. 02640232)(A.H.).

\vskip 1 cm

\newpage
\noindent {\bf Figure captions}
\bigskip

\noindent $\bullet$ {\bf Figure 1.}
The spectrum of the adiabatic baryonic perturbation.
The dashed and solid lines represent the initial scale invariant
spectrum and the spectrum at the recombination time, respectively.
$l_c$ is the Jeans length at the equal time.

\noindent $\bullet$ {\bf Figure 2.}
The diagram shows the time evolution of some characteristic lengths,
the Hubble horizon, the Jeans length at equal time $l_c$, the mean
free path $l_{mf}$, and the localization length $\xi$.

\noindent $\bullet$ {\bf Figure 3.}
The diffusion of the localized electromagnetic field.
The upper figure represents the localized field before the
recombination time, while the lower one represents the diffusion process
after that time.

\end{document}